\documentclass[final,a4paper,openright,twoside,11pt]{article}
\usepackage{amssymb,amsmath,latexsym}
\usepackage{graphicx}
\usepackage{color}
\usepackage[usenames,dvipsnames]{xcolor}
\usepackage[utf8]{inputenc}
\usepackage[english]{babel}
\usepackage[]{natbib}
\usepackage{pgfgantt}
\usepackage{epstopdf}
\usepackage{multirow}
\usepackage{rotating}
\usepackage{pdflscape}
\usepackage{amsmath}
\usepackage{enumitem}  
\usepackage{hyperref}

\makeatletter
\def\@seccntformat#1{%
	\expandafter\ifx\csname c@#1\endcsname\c@section\else
	\csname the#1\endcsname\quad
	\fi}
\makeatother

\begin{document}
\pagenumbering{gobble}
\setcounter{page}{1}

\title{German to Spanish translation of Einstein’s work on the formation of meanders in rivers.}

\author{Enrique M. Padilla\thanks{Enrique M. Padilla and Manuel Díez-Minguito are with the Andalusian Institute for Earth System Research (IISTA), University of Granada, Spain (EU). (email: epadilla@ugr.es)} , Birgit L. Emberger, Manuel Díez-Minguito${^*}$
}

\date{}

\maketitle

\begin{abstract}
{In 1926 Albert Einstein gave a clear explanation of the physical processes involved in the meander formation and evolution in open channels \citep{Einstein1926}. Although this work is far from being recognized as one of his greatest achievements, such as his \textit{annus mirabilis} papers in 1905, he shows a truly remarkable didactic skills that make it easy to understand even to the non-specialist. In particular, a brilliant explanation of the tea leaf paradox can be found in this paper of 1926, presented as a simple experiment for clarifying the role of Earth rotation and flow curvature in the differential river banks erosion. This work deserves to be considered as a pioneering work that has laid a basic knowledge in currently very active research fields in fluvial geomorphology, estuarine physics, and hydraulic engineering. In response to the curiosity aroused and transmitted to the authors over the years by undergraduates and MSc. students, and also due to its historical and scientific significance, we present here the Spanish translation of Einstein’s original work published in German in 1926 in \textit{Die Naturwissenschaften} \citep{Einstein1926}. Einstein’s drawings have not been interpreted, but just updated preserving their original spirit.}
\end{abstract}

\pagenumbering{arabic}
\setcounter{page}{1}

\vspace*{0.3cm}
\rule{130mm}{0.1mm}

\section{La causa de la formación de meandros en los cursos de los ríos y de la así llamada ley de Baer.}

{\large Por Albert Einstein, Berlín.}
\\
\\

\begin{figure}[t!]	
	\begin{center}
		\includegraphics[width=0.7\textwidth]{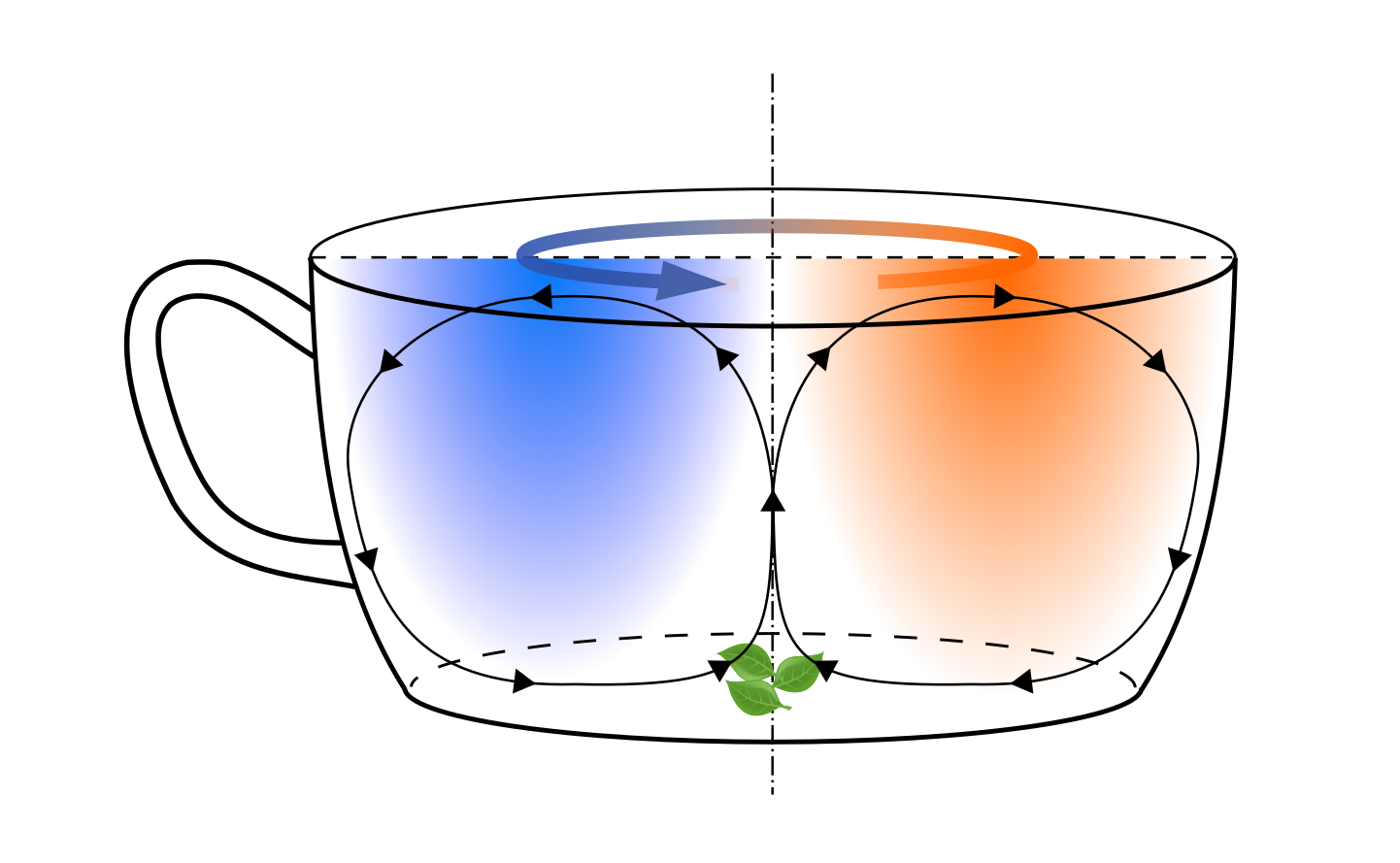}
		\caption{Actualizada de \cite{Einstein1926}, Fig 1.}
		\label{Fig1}
	\end{center}	
\end{figure}

Es de común conocimiento que las corrientes de los ríos tienden a curvarse y a serpentear en vez de seguir la línea de máxima pendiente del terreno. También es bien sabido por los geógrafos que los ríos del hemisferio norte tienden a erosionar principalmente en la margen derecha. Los ríos del hemisferio sur se comportan de manera opuesta (Ley de Baer). Se han hecho muchos intentos por explicar este fenómeno, y no estoy seguro de si algo de lo que diga a continuación aún sea novedoso para el experto; algunas de las consideraciones que se expondrán son ciertamente conocidas. No obstante, al no haber encontrado a nadie que estuviera completamente familiarizado con las relaciones causales involucradas, creo que es apropiado dar a continuación una breve exposición cualitativa de ellas.
En primer lugar, está claro que más fuerte debe ser la erosión cuanto mayor sea la velocidad de la corriente cerca de la orilla en cuestión, o cuanto más rápido decaiga la corriente a cero en un punto concreto de la margen del río. Esto es igualmente cierto en todas las circunstancias, ya dependa la erosión de factores mecánicos o fisicoquímicos (descomposición del suelo). Debemos pues centrar nuestra atención en las circunstancias que afectan a la inclinación del gradiente de velocidad en la pared.
En ambos casos, la asimetría en cuanto a la caída de la velocidad en cuestión se debe indirectamente a la formación de un movimiento circular al que a continuación dirigiremos nuestra atención. Comienzo con un pequeño experimento que todo el mundo puede repetir fácilmente. Imagine una taza de fondo plano llena de té. 
En la parte inferior debe haber algunas hojas de té, que permanecen ahí por ser bastante más pesadas que el líquido que han desalojado. Si se hace girar el líquido con una cuchara, las hojas pronto se acumularán en el centro del fondo de la taza. La explicación de este fenómeno es la siguiente: la rotación del líquido provoca una fuerza centrífuga que actúa sobre él. Esta en sí misma no daría lugar a ningún cambio en el flujo del líquido si este último girara como un cuerpo rígido. Pero en las proximidades de las paredes de la taza, el movimiento del líquido es retenido por la fricción, de modo que circula allí con menor velocidad angular que en otros lugares más interiores. En particular, la velocidad angular de rotación, y por lo tanto la fuerza centrífuga, será menor cerca del fondo que a mayor altura. Esto daría lugar a la formación de un movimiento circular del líquido del tipo ilustrado en la Figura \ref{Fig1} que se va incrementando hasta que, bajo la influencia de la fricción del fondo, se haga estacionaria. Las hojas de té son barridas hacia el centro por el movimiento circular y actúan como prueba de su existencia. 

\begin{figure}[t!]	
	\begin{center}
		\includegraphics[width=0.7\textwidth]{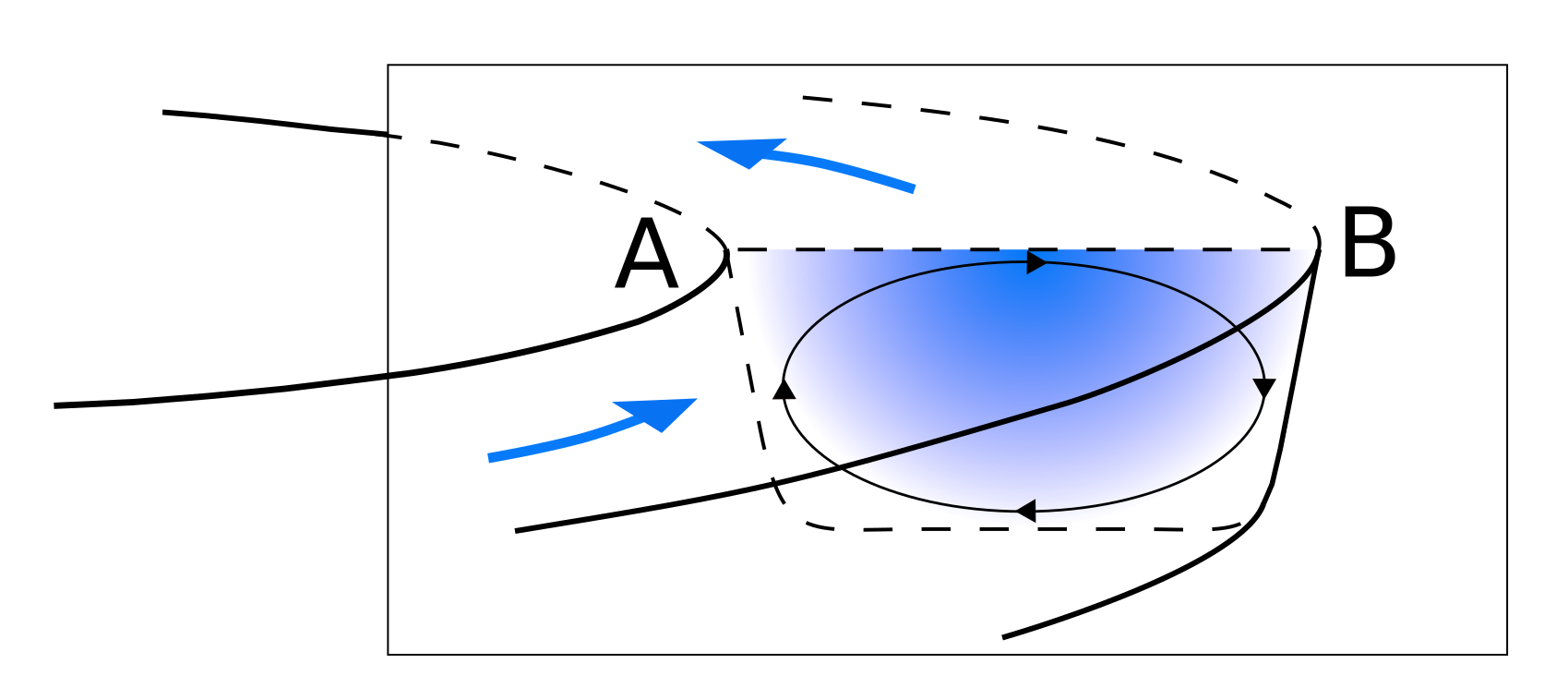}
		\caption{Actualizada de \cite{Einstein1926}, Fig 2.}
		\label{Fig2}
	\end{center}	
\end{figure}

La situación es análoga a la de un río que se curva (Figura \ref{Fig2}). En cada sección transversal de su curso, donde esté curvado, actúa una fuerza centrífuga dirigida hacia el exterior de la curva (de A a B). Esta fuerza, sin embargo, en el fondo, donde la velocidad de la corriente se reduce por el efecto de la fricción, es menor que a mayor altura. Así se forma una circulación del tipo ilustrado en la Figura \ref{Fig2}. Pero incluso donde no haya ninguna curva en el río, se desarrollará un movimiento circular del tipo mostrado en la Figura \ref{Fig2}, aunque a pequeña escala, como resultado de la rotación de la Tierra. Ésta última produce una fuerza de Coriolis que actúa transversalmente a la dirección de la corriente, cuya componente horizontal a la derecha es $2\,\upsilon\,\Omega \sin(\varphi)$  por unidad de masa del líquido, donde $\upsilon$ es la velocidad de la corriente, $\Omega$ la velocidad de la rotación de la Tierra y $\varphi$ la latitud geográfica. Como la fricción del suelo causa una disminución de esta fuerza hacia el fondo, esta fuerza también da lugar a un movimiento circular del tipo indicado en la Figura \ref{Fig2}.

Después de esta discusión preliminar volvemos ahora a la cuestión de la distribución de las velocidades en la sección transversal de la corriente, que es el factor que controla la erosión. Para ello, primero debemos considerar cómo la distribución de velocidades (turbulenta) en un río se desarrolla y es mantenida. Si el agua que estaba previamente en reposo fuera puesta repentinamente en movimiento por la acción de una fuerza de aceleración distribuida uniformemente, entonces la distribución de velocidades en la sección transversal sería en un primer momento uniforme. Sólo después de un tiempo bajo la influencia de la fricción en las paredes se establecería una distribución de las velocidades en la que estas aumentaran gradualmente desde las paredes hacia el centro de la sección transversal. Una perturbación de esta (grosso modo) distribución estacionaria de velocidades en la sección transversal volvería a equilibrarse (lentamente bajo la influencia de la fricción del fluido). La hidrodinámica describe el proceso mediante el cual se establece esta distribución estacionaria de velocidades de la siguiente manera. En una distribución de flujo uniforme (flujo potencial) todos los filamentos de vórtices se concentran en las paredes. Se desprenden y lentamente se desplazan hacia el interior de la sección transversal de la velocidad, distribuyéndose en una capa de espesor creciente. De este modo, el gradiente de velocidad en las paredes disminuye lentamente. Bajo la acción de la fricción interna del fluido, los filamentos de vórtices en el interior de la sección transversal son lentamente absorbidos, siendo reemplazados por otros nuevos que se forman en la pared. De esta manera se produce una distribución cuasi-estacionaria de velocidades. Lo importante para nosotros es que la igualación de la distribución de velocidades a una distribución estacionaria de velocidades es un proceso lento. A ello se debe que incluso causas relativamente insignificantes, pero de acción constante, sean capaces de ejercer una influencia considerable en la distribución de velocidades en la sección transversal. 

Consideremos ahora qué tipo de influencia debe ejercer el movimiento circulatorio causado por una curva en el río o la fuerza de Coriolis, ilustrado en la Figura \ref{Fig2}, en la distribución de velocidades sobre la sección transversal del río. Las partículas de líquido más aceleradas serán las más alejadas de las paredes, es decir, se hallarán en la parte superior de la parte central del fondo. Estas partes, las más rápidas del agua, serán impulsadas por la circulación hacia la pared del lado derecho, mientras que por el contrario la pared del lado izquierdo recibe el agua que proviene de la región cercana al fondo con una velocidad especialmente baja. Por consiguiente (en el caso de la Figura \ref{Fig2}) la erosión es necesariamente más intensa sobre el lado derecho que sobre el izquierdo. Cabe señalar que esta explicación está basada esencialmente en el hecho de que el lento movimiento circulatorio del agua ejerce una influencia considerable en la distribución de velocidades, ya que el proceso de equilibrio de las velocidades por fricción interna, la cual contrarresta esta consecuencia del movimiento circulatorio, es también un proceso lento. 

\begin{figure}[t!]	
	\begin{center}
		\includegraphics[width=0.7\textwidth]{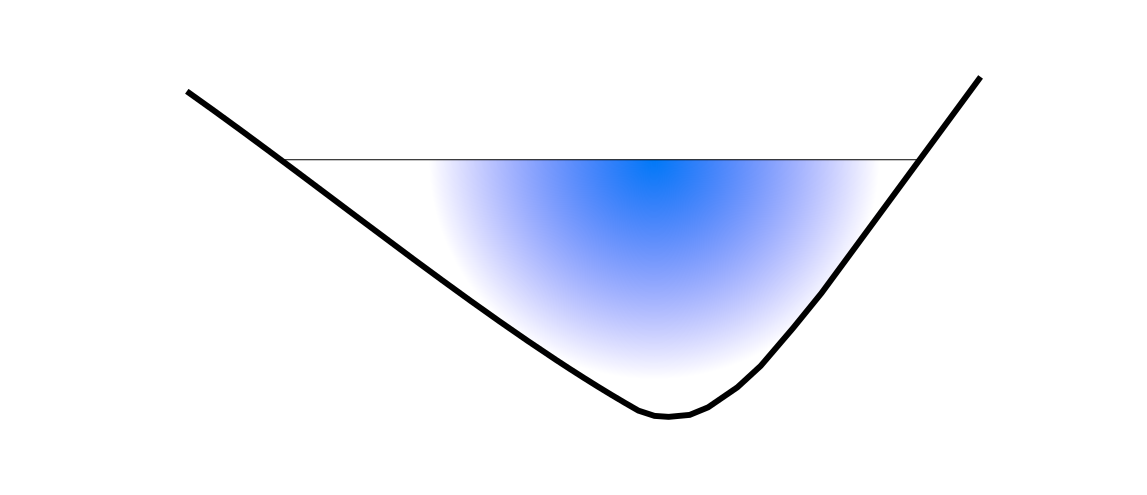}
		\caption{Actualizada de \cite{Einstein1926}, Fig 3.}
		\label{Fig3}
		\end{center}	
\end{figure}

Con esto hemos dilucidado las causas de la formación de meandros. Pero también ciertos detalles pueden deducirse fácilmente de estos hechos. La erosión tendrá que ser relativamente grande no solo en la pared lateral derecha, sino también en la parte derecha del fondo, de modo que habrá una tendencia a formar un perfil de la forma indicada en la Figura \ref{Fig3}. Además, el agua en la superficie vendrá de la pared lateral izquierda, y por lo tanto (especialmente en el lado izquierdo) se moverá menos rápidamente que el agua a una profundidad algo mayor. Esto, en efecto, se ha observado. Asimismo, hay que tener en cuenta que el movimiento circulatorio posee inercia. La circulación, por tanto, alcanzará su valor máximo sólo corriente abajo del lugar de mayor curvatura, y lo mismo se aplica naturalmente a la asimetría de la erosión. Así, el proceso erosivo irá acompañado por un avance de las líneas de ondas de la formación del meandro en el sentido de la corriente. Por último, cuanto mayor sea la sección transversal del río, tanto más lentamente el movimiento circulatorio será absorbido por la fricción; así pues, la longitud de onda de la formación de los meandros aumentará con el tamaño de la sección transversal del río.


\end{document}